\documentstyle[12pt]{article}
\newcommand{\dfrac}[2]{{\displaystyle{\frac{#1}{#2}}}}

\newcommand{\journal}[4]{{\em #1~} {\bf #2} (19#3) #4;}

\newcommand{\np}{\journal {Nucl. Phys.}}
\newcommand{\pl}{\journal {Phys. Lett.}}

\newcommand{\G}{\Gamma}
\newcommand{\D}{\Delta}
\renewcommand{\a}{\alpha}
\renewcommand{\b}{\beta}
\renewcommand{\d}{\delta}
\newcommand{\e}{\varepsilon}
\newcommand{\f}{\phi}
\newcommand{\g}{\gamma}

\newcommand{\x}{\xi}
\renewcommand{\l}{\lambda} \renewcommand{\L}{\Lambda}
\newcommand{\m}{\mu}
\newcommand{\n}{\nu}
\newcommand{\mn}{{\mu\nu}}

 \renewcommand{\O}{\Omega}
\newcommand{\p}{\psi}
\renewcommand{\r}{\rho}
\newcommand{\s}{\sigma} \renewcommand{\S}{\Sigma}
\newcommand{\th}{\theta}

\renewcommand{\t}{\tau}

\renewcommand{\AA}{{\cal A}}

\newcommand{\CC}{{\cal C}}

\newcommand{\GG}{{\cal G}}
\newcommand{\HH}{{\cal H}}
\newcommand{\II}{{\cal I}}

\newcommand{\KK}{{\cal K}}

\newcommand{\MM}{{\cal M}}
\newcommand{\NN}{{\cal N}}
\newcommand{\OO}{{\cal O}}
\newcommand{\PP}{{\cal P}}

\newcommand{\RR}{{\cal R}}
\newcommand{\SS}{{\cal S}}

\newcommand{\VV}{{\cal V}}

\newcommand{\complex}{{\kern .1em {\raise .47ex
\hbox {$\scriptscriptstyle |$}}
    \kern -.4em {\rm C}}}
\newcommand{\real}{{{\rm I} \kern -.19em {\rm R}}}
\newcommand{\rational}{{\kern .1em {\raise .47ex
\hbox{$\scripscriptstyle |$}}
    \kern -.35em {\rm Q}}}

\newcommand{\cb}{{\bar c}}

\newcommand{\pa}{\partial}

\newcommand{\sla}{\raise.15ex\hbox{$/$}\kern -.57em}

\newcommand{\twiddle}{\lower.9ex\rlap{$\kern -.1em\scriptstyle\sim$}}

\newcommand{\dxm}{dx^\mu }
\newcommand{\dxn}{dx^\nu }

\newcommand{\vf}{{\varphi}}
\newcommand{\pam}{{\partial_\mu}}

\newcommand{\Ot}{{\tilde\Omega}}

\newcommand{\eq}{\begin{equation}}
\newcommand{\eqn}[1]{\label{#1}\end{equation}}
\newcommand{\eea}{\end{eqnarray}}
\newcommand{\eqa}{\begin{eqnarray}}
\newcommand{\eqan}[1]{\label{#1}\end{eqnarray}}
\newcommand{\ba}{\begin{array}}
\newcommand{\ea}{\end{array}}
\newcommand{\eqac}{\begin{equation}\begin{array}{rcl}}
\newcommand{\eqacn}[1]{\end{array}\label{#1}\end{equation}}

\def\non{\nonumber\\}

\def\cb{\bar{c}}
\def\lb{\bar{\l}}
\def\6{\partial}
\def\={\!\!\!&=&\!\!\!}
\def\+{\!\!\!&&\!\!\!+~}
\def\-{\!\!\!&&\!\!\!-~}

\def\ve{\varepsilon}

\renewcommand{\title}[1]{\null\vspace{25mm}

\noindent{\Large{\bf #1}}\vspace{10mm}

\noindent {\large  }}
\newcommand{\authors}[1]{\noindent{\large #1}\vspace{3mm}}

\newcommand{\address}[1]{\noindent #1\vspace{5mm}

}
\renewcommand{\abstract}[1]{\vspace{10mm}

\noindent{\small{\em Abstract.} #1}\vspace{2mm}

} 

\begin{document}

\setcounter{page}{1}
\thispagestyle{empty}\hspace*{\fill}REF. TUW 98-02
\begin{center}
\title{Finiteness of 2D Topological BF-Theory \\
        with Matter Coupling\\}
\authors{R. Leitgeb\footnote{Work supported in part by the
``Fonds zur F\"orderung der Wissenschaftlichen Forschung'',
under\\[1pt] Contract Grant Number P11354 -- PHY, and by the
``Magistrat der Stadt Wien, MA18''
under Contract\\[1pt] Grant Number MA18 -- WI/342/98.},
  M. Schweda and 
H. Zerrouki\footnote{Work supported in part by the
``Fonds zur F\"orderung der Wissenschaftlichen Forschung'',
under Contract Grant Number P11582 -- PHY.}}

\address{Institut f\"ur Theoretische Physik,
         Technische Universit\"at Wien,\\
         Wiedner Hauptstra\ss e 8-10,
         A-1040 Wien, Austria} 
\end{center}

\abstract{ 
We study the ultraviolet and the infrared behavior
of $2D$ topological BF-Theory coupled to vector and scalar fields.
This model is equivalent to $2D$ gravity coupled to topological matter.
Using techniques of the algebraic renormalization program we
show that this model is anomaly free and ultraviolet
as well as infrared finite at all orders of perturbation theory.
}

\newpage

\section{Introduction}

During the last decade the study of topological gauge theories provided
deep insights into the topology and geometry of low dimensional manifolds.
The central feature of topological field theories is that the observables 
only depend on the global structure of the space-time manifold on which the
model is defined. 
There are in fact two different 
types of topological field theories: whether the whole gauge fixed action 
or just the gauge fixing part can be  written as a BRS variation the model is 
of Witten-type or of Schwarz-type, respectively. An example of 
Witten-type models is the topological Yang-Mills theory, representatives of
Schwarz-type models are Chern-Simons and BF theories. A common feature of
many such field theories is the appearance of a so-called vector-like 
supersymmetry in the flat space-time limit. It is due to the energy-momentum
tensor being BRS-exact. Its generators when anticommuted with the BRS operator 
close on translations. 

We are in particular interested in a BF-model defined on a two dimensional 
space-time manifold. Such a model was shown to be         
equivalent to a two dimensional gravity, which has
been discussed in connection with the Liouville theory \cite{2dgravity}. 
Typically in two space-time dimensions the propagators of massless scalar 
fields are ill-defined \cite{balasin, olivier2}.
Due to the singular behaviour of the ghost propagator at long distances,
an infrared regulator mass has to be introduced \cite{blasi}. The infrared
and ultraviolet finiteness of the two dimensional BF-model was
already discussed in the realm of algebraic renormalization \cite{blasi}. 

The present work is devoted to the investigation of an enlarged model with
the inclusion of a topological matter interaction \cite{2dgravity} in the 
context
of the algebraic renormalization program \cite{olivier2}.
The resulting model is characterized by an enlarged BRS symmetry. 
Moreover, we show the existence of a  vector-like supersymmetry, 
which fact simplifies the investigation of the infrared 
and ultraviolet renormalizability of this model.

The paper is organized as follows. 
In section 2 we describe the model at the classical level and we display its 
BRS transformations as well as the vector-like supersymmetry transformations.
In section 3 we prove the finiteness of the model.
Finally, in section 4, we show that the model is anomaly free. 

\section{Definition of the model at the classical level}

Let us first consider the BF model on a two dimensional flat
manifold $\MM$ endowed with an Euclidean metric $\eta_\mn$.
This field model possesses the following metric independent action:
\eq
\S^{(1)}_{inv}=\frac{1}{2}\int_{\MM}d^2x~ \e^{\m\n}F^a_{\m\n}\f^a\,,
\eqn{action1}
where $\e^{\m\n}$ is the completely antisymmetric Levi--Civita tensor 
(with $\e^{12}=+1$), $\f^a$ is a scalar field, and the 
field strength $F^a_{\m\n}$ is given by
\eq
F^a_{\m\n}=\6_\m A^a_\n -\6_\n A^a_\m + f^{abc}A^b_\m A^c_\n\,.
\eqn{8345ljkdfg}
$A_\m^a$ stands for the gauge field with gauge index $a$.
$f^{abc}$ are the structure constants of the gauge group
which is assumed to be a compact and simple Lie group with all
fields belonging to its adjoint representation. 
The generators of the Lie algebra are chosen to be anti-hermitian, 
such that $[T^a,T^b]=f^{abc} T^c$ and $Tr(T^a T^b)= \d^{ab}$.\\  
The action $\S^{(1)}_{inv}$ is invariant under the following infinitesimal
gauge transformations
\eqa
\d_{(\th)} A^a_\m &=& (\pam \th^a + f^{abc} A^b_\m\th^c) \equiv (D_\m\th)^a\,, 
\nonumber\\
\d_{(\th)} \f^a &=& -f^{abc}\th^b\f^c\,,
\eqan{gauge1}
where $\th^a$ is a local parameter, and $D_\m$ is the covariant derivative.
This model has already been studied in much detail \cite{blasi}. In the present
work we enlarge the model by introducing a set of $N$ vector fields 
$B_\m^{a\a}$ and $N$ scalars $X^{a\a}$ \cite{2dgravity}, 
where the index $\a$ takes all values from 1 to $N$.
The contribution of these new fields represents a matter interaction. It is 
given by the metric independent local functional
\eq
\S_{N}=\int_\MM d^2x ~ \e^\mn(D_\m B^\a_\n)^a X^a_\a \,.
\eqn{sdkfjgh}
Without loss of generality, we restrict ourselves to the case $N=1$
implying that (\ref{sdkfjgh}) now reads
\eq
\S^{(2)}_{inv}=\int_\MM d^2x ~ \e^\mn(D_\m B_\n)^a X^{a} \,.
\eqn{action2}
The full action $\S_{inv}=\S^{(1)}_{inv}+
\S^{(2)}_{inv}$ is in fact
invariant under an additional symmetry \cite{2dgravity} given by the 
transformations
\eqa
\d_{(\p)}\f^a&=&-f^{abc}\p^b X^{c}\,,\nonumber\\
\d_{(\p)} B_\m^{a}&=&D_\m\p^{a}\,,
\eqan{gauge2} 
where $\p^a$ is an infinitesimal parameter.\\
As usual, the quantisation
procedure requires a gauge fixing. In the present case we 
use a Landau-type gauge. Due to the fact that we have two gauge symmetries we
need two sets of ghost fields with corresponding Lagrange multiplier fields:
($c^a$,$\l^{a}$) are the Faddeev-Popov ghost fields with 
corresponding
antighost fields ($\cb^a$,$\lb^{a}$), and ($b^a$,$d^{a}$) are 
Lagrange multiplier fields enforcing the Landau gauge conditions.
The gauge fixed action is then given by
$\Sigma_{inv}+\Sigma_{gf}$, where 
\eqa
\S_{gf}&=&s\int_\MM d^2x (\cb^a\6_\m A^{a\m }+ 
\lb^{a}\6_\m B^{a\m})=\nonumber\\
 &=&\int_\MM d^2x[b^a\6_\m A^{a\m}+d^{a}\6_\m B^{a\m}]+
\nonumber\\
 &+& \int_\MM d^2x [-\cb^a \6_\m (D^\m c)^a -\lb^{a} \6_\m (D^\m \l)^a +
f^{abc} \lb^{a} \6_\m(c^b B^{c\m})]\,.
\eqan{actiongf}
The next step is to promote the two gauge symmetries to the nilpotent and 
nonlinear BRS-symmetry of the gauge fixed action 
\eqa
sA_\m^a&=&(D_\m c)^a\,,\nonumber\\ 
sB_\m^{a}&=& (D_\m \l)^{a}- f^{abc} c^b B_\m^{c}\,,\nonumber\\ 
s\f^a &=& -f^{abc} (c^b \f^c+\l^{b} X^c)\,,\nonumber\\
sX^{a}&=& -f^{abc} c^b X^{c}\,,\nonumber\\
sc^a&=&-\frac{1}{2}f^{abc} c^b c^c\,,\nonumber\\
s\l^{a}&=&-f^{abc} c^b \l^{c}\,,\nonumber\\
s\cb^a&=& b^a\,,\qquad sb^a=0\,,\nonumber\\
s\lb^{a}&=& d^{a}\,,\qquad sd^{a}=0\,,\nonumber\\
s^2&=&0\,.
\eqan{brst}
The canonical dimensions and Faddeev-Popov charges of all fields introduced so
far are listed in (Table \ref{fp1}).
\begin{table}[h]
\begin{center} \begin{tabular}{|l|r|r|r|r|r|r|r|r|r|r|r|} \hline
 & $A^a_\m$ & $B^a_\m$ & $\f^a$ & $X^a$ & $c^a$ & $\l^a$ & $\cb^a$ & $\lb^a$ 
& $b^a$ & $d^a$ & $\6_\m$ \\ \hline
 dim & 1 & 1 & 0 & 0 & 0 & 0 & 0 & 0 & 0 & 0 & 1 \\ \hline
$\f\pi$ & 0 & 0 & 0 & 0 & 1 & 1 & -1 & -1 & 0 & 0 & 0\\ \hline
\end{tabular}\end{center}\caption{Dimensions and Faddeev-Popov charges of the
fields.}\label{fp1}
\end{table}
\\    
Note, that the BRS-exact gauge fixing term introduces a metric. As a 
consequence the energy-momentum tensor is BRS-exact as well and the model 
possesses a further symmetry carrying a vectorial index:
\eq
\begin{array}{l@{$\,=\,$}ll@{$\,=\,$}l}
\d_\m A_\n^a & 0\,, & \d_\m B_\n^{a} & 0\,, \\ 
\d_\m \f^a & -\e_\mn \6^\n\cb^a\,, & \d_\m X^{a} & -\e_\mn \6^\n\lb^{a}\,,\\
\d_\m c^a & A_\m^a \,, & \d_\m \l^{a} & B_\m^{a}\,,\\
\d_\m \cb^a & 0\,, & \d_\m \lb^{a} & 0 \,,\\
\d_\m b^a & \6_\m\cb^a\,, & \d_\m d^{a} & \6_\m\lb^{a} \,.\\
\end{array}
\eqn{vsusy}
The transformations (\ref{vsusy}) define the so-called vector-like 
supersymmetry
\footnote{We will see in due course that the algebra of $\d_\m$ with the 
BRS operator
$s$ closes on translations.} 
\cite{oli}.
The invariant action plus the gauge fixing part is indeed invariant 
under (\ref{vsusy}),
\eq \d_\m (\S_{inv}+\S_{gf})=0\,. \eqn{invar}   
Additionally, the symmetries (\ref{brst}) and (\ref{vsusy}) give rise to the 
following on-shell algebra: 
\eqa
\{s,s\} &=& 0\,,\nonumber\\
{\{s,\d_\m\}} &=& \6_\m + \hbox{equations of motion}\,,\nonumber\\
{\{\d_\m,\d_\n\}} &=& 0\,.
\eqan{algebra}
In order to write down the Slavnov identity, which expresses the symmetry 
content of the model with respect to BRS, we couple the nonlinear 
BRS transformations to BRS-invariant external sources leading to 
\eq
\S_{ext}=\int_\MM d^2x[\O^{\m a}(s A_\m^a)+L^a(sc^a)+\r^a(s\f^a)+
\s^{\m a}(sB^a_{\m})+\L^{a}(s\l^a)+Y^{a}(sX^a)]\,,
\eqn{actionext}
where $(\O^{\m a},L^a,\r^a,\s^{\m a},\L^{a},Y^{a})$ are the external 
sources
\footnote{Clearly, the presence of these external sources breaks the symmetry 
(\ref{vsusy}). This fact will be discussed later.}
whose canonical dimensions and Faddeev-Popov charges can be read 
off (Table \ref{fp2}).\\
\begin{table}[h]
\begin{center} \begin{tabular}{|l|r|r|r|r|r|r|} \hline
 & $\O^{\m a}$  & $L^a$  & $\r^a$ & $\s^{\m a}$ & $\L^{a}$ & $Y^{a}$\\ 
\hline
 dim & 1 & 2 & 2 & 1 & 2 & 2 \\ \hline
$\f\pi$ & -1 & -2 & -1 & -1 & -2 & -1  \\ \hline
\end{tabular}\end{center}
\caption{Dimensions and Faddeev-Popov charges of the
external sources.}
\label{fp2}
\end{table}

Typically the propagators of massless scalar fields are ill-defined in two
space-time dimensions \cite{balasin, olivier2}. In particular, the analysis of the infrared problem 
connected with the ghost-antighost
propagator
\eq
\langle \cb^a c^b \rangle = \d^{ab} \frac{1}{k^2}\,, 
\eqn{x12x}
requires the introduction of a regulator mass $m$ \cite{blasi} such that
\eq
\langle \cb^a c^b \rangle_m = \d^{ab} \frac{1}{k^2+m^2}\,.
\eqn{propc}
Keeping the algebraic structure (\ref{algebra}) amounts to 
adding\footnote{In fact, it would be sufficient to add to the action 
the quantity
\eqa
\S_\tau &=& \int_\MM d^2x \left ( -(\t_1+m^2)\cb^a c^a - \t_2(b^ac^a+
\frac{1}{2}f^{abc} \cb^a c^b c^c) \right) \non
&=& s\int_\MM d^2x(\t_2 \cb^a c^a )\,, \nonumber
\eqan{exp4laining}
which guarantees (\ref{propc}) and the BRS
invariance of the total action. However, 
to maintain also the vector supersymmetry 
(allowing only linear breaking terms in the
corresponding  Ward identity as well as keeping the algebraic structure
(\ref{algebra}), see below) we need the additional expression
$\int_\MM d^2x \ s(\tau^\m_4 \bar c^a A_\m^a)$ 
containing the external sources $\tau^\m_3$ and $\tau^\m_4$.} 
the following expression \cite{blasi} to the action
\eqa
\S_m&=& \int_\MM d^2x \left ( -(\t_1+m^2)\cb^a c^a - \t_2(b^ac^a+
\frac{1}{2}f^{abc} \cb^a c^b c^c) + \t_3^\m \cb^a A_\m^a + 
\right.\nonumber \\
 &+& \left.\frac{}{}\t_4^\m(b^a A_\m^a - \cb^a(D_\m c)^a)\right)=\nonumber\\
 &=& s\int_\MM d^2x[\t_2 \cb^a c^a + \t_4^\m \cb^a A_\m^a]\,, 
\eqan{actionm}
The quantities $(\t_1,\t_2,\t_3^\m,\t_4^\m)$ 
are the new external sources with the following BRS transformation laws
\eq
\ba{ll}
s\t_2= -(\t_1+m^2), ~~~~&~~~~ s \t_1=0, \\
s\t_4^\m=\t_3^\m,   ~~~~&~~~~ s \t_3^\m = 0\,, \\
\ea
\eqn{brsextm}
and with dimensions and Faddeev Popov charges as given in (Table \ref{fp3}).
\begin{table}[h]
\begin{center} \begin{tabular}{|l|c|c|c|c|} \hline
 & $\t_1$  & $\t_2$  & $\t_3^\m$ & $\t_4^\m$\\ \hline
 dim & 2 & 2 & 1 & 1 \\ \hline
$\f\pi$ & 0 & -1 & 1 & 0  \\ \hline
\end{tabular}
\end{center}
\caption{Dimensions and Faddeev-Popov charges of the
external sources $\tau$.}
\label{fp3}
\end{table}
\noindent

The same strategy can be used to regularize the 
propagator
\eq
\langle \lb^{a} \l^b \rangle = \d^{ab} \frac{1}{k^2}\,.
\eqn{yxy}
Correspondingly, we have to add the following BRS exact expression to 
the action
\eqa
\S_m'&=& \int_\MM d^2x \left ( - ( \tau_{1}+m^2) \lb^{a}\l^a
 - \tau_2(d^{a} \l^{a}+
f^{abc} \lb^{a} c^b \l^c) + \tau_3^\m \lb^{a} B_\m^a + 
\right.\nonumber \\
 &+& \left.\frac{}{}\tau_4^\m[d^{a} B_\m^a - \lb^{a}
(D_\m \l)^a+
f^{abc}\lb^{a} c^b B_\m^c]\right)\nonumber\\
 &=& s\int_\MM d^2x[\tau_2\lb^{a} \l^{a} + \tau_4^\m\lb^{a} B_\m^{a}]\, .
\eqan{actionM}
It follows that the propagator (\ref{yxy}) is regularized according to 
\eq
\langle \lb^{a} \l^b \rangle_m = \d^{ab} \frac{1}
{k^2+ m^2}\,.
\eqn{propl}
The total action, which is just the vertex functional at the classical
level, becomes
\eq
\S^{(0)}=\S_{inv}+\S_{gf}+ \S_{ext} + \S_{\II \RR},
\eqn{totaction}
with
\eq
\S_{\II \RR}= \S_m + \S_m' = s \int_\MM d^2x \left( \t_2 (\cb^ac^a + 
              \bar \l^a \l^a) + \t_4^\m(\cb^a A^a_\m + \bar \l^a
              B^a_\m) \right).
\eqn{ljf898}  
We are now ready to write down the non-linear functional Slavnov identity 
corresponding 
to the BRS invariance of the total action
\eqa
\SS(\S^{(0)}) &=& \int_\MM d^2x \left(
          \frac{\d\S^{(0)}}{\d\O^{a\m}}\frac{\d\S^{(0)}}{\d A_\m^a}+
          \frac{\d\S^{(0)}}{\d L^a}\frac{\d\S^{(0)}}{\d c^a }+
          \frac{\d\S^{(0)}}{\d\r^a}\frac{\d\S^{(0)}}{\d\f^a}+
          \frac{\d\S^{(0)}}{\d\s^{a\m}}\frac{\d\S^{(0)}}{\d B^a_{\m}}+
          \frac{\d\S^{(0)}}{\d\L^{a}}\frac{\d\S^{(0)}}{\d\l^a}+ \right.  
          \nonumber \\ &+& 
          \left. \frac{\d\S^{(0)}}{\d Y^{a}}\frac{\d\S^{(0)}}{\d X^a}+
          b^a \frac{\d\S^{(0)}}{\d \cb^{a}} + 
          d^a \frac{\d\S^{(0)}}{\d \lb^{a}} 
          -(\t_1+m^2)\frac{\d \S^{(0)}}{\d\t_2}+
          \t_3^\m\frac{\d \S^{(0)}}{\d\t_4^\m} \right)=0\,.
\eqan{slavnovid}
For later use, we derive from (\ref{slavnovid}) the linearized version 
$\SS_{\S^{(0)}}$
of the non-linear BRS-operator:
\eqa
{\SS_{\S^{(0)}}} &=& \int_\MM d^2x \left(\frac{\d\S^{(0)}}{\d\O^{a\m}}
\frac{\d}{\d A_\m^a}+
\frac{\d\S^{(0)}}{\d A_\m^a}\frac{\d}{\d\O^{a\m}}+
\frac{\d\S^{(0)}}{\d L^a}\frac{\d}{\d c^a }+
\frac{\d\S^{(0)}}{\d c^a }\frac{\d}{\d L^a}+
\frac{\d\S^{(0)}}{\d\r^a}\frac{\d}{\d\f^a}+
\frac{\d\S^{(0)}}{\d\f^a}\frac{\d}{\d\r^a}+
\right. \nonumber\\
 &+& 
\frac{\d\S^{(0)}}{\d\s^{a\m}}\frac{\d}{\d B^a_{\m}}+
\frac{\d\S^{(0)}}{\d B^a_{\m}}\frac{\d}{\d\s^{a\m}}+
\frac{\d\S^{(0)}}{\d\L^{a}}\frac{\d}{\d\l^a}+
\frac{\d\S^{(0)}}{\d\l^a}\frac{\d}{\d\L^{a}}+
\frac{\d\S^{(0)}}{\d Y^{a}}\frac{\d}{\d X^a}+
\frac{\d\S^{(0)}}{\d X^a}\frac{\d}{\d Y^{a}}+
\nonumber\\ &+& \left. 
b^a \frac{\d}{\d \cb^{a}} + d^{a} \frac{\d}{\d \lb^{a}} 
-(\t_1+m^2)\frac{\d}{\d\t_2}+\t_3^\m\frac{\d}{\d\t_4^\m}
\right)\,. 
\eqan{slavnovop}  
As already mentioned, the introduction of external sources breaks the vector-like
supersymmetry. This fact is expressed by the following broken Ward-identity
(WI) for the symmetry (\ref{vsusy}):
\eq
\VV_\m\S^{(0)} = \D_\m^{cl}\,,
\eqn{invar2}
where ${\VV_\m}$ is the vector--like supersymmetry
Ward operator 
\eqa
\VV_\m &=& \int_\MM d^2x \Bigg( \e_\mn\r^a \frac{\d}{\d A_\n^a}
-\e_\mn[\O^{\n a}+\6^\n\cb^a -\t_4^\n\cb^a]\frac{\d}{\d\f^a}
+A_\m^a\frac{\d}{\d c^a }+(\6_\m \cb^a)\frac{\d}{\d b^a}
+ L^a \frac{\d}{\d \O^{\m a}}
\nonumber \\
&+& B_\m^a \frac{\d}{\d \l^a}
+\e_\mn Y^a \frac{\d}{\d B^a_\n}-
\e_\mn [\s^{\n a}+\6^\n\lb^a -\t_4^\n\lb^a]\frac{\d}{\d X^a}
+(\6_\m \lb^a)\frac{\d}{\d d^a}+
\L^a \frac{\d}{\d \s^{\m a}}
\nonumber\\
&-& (\6_\m\t_2)\frac{\d}{\d\t_1}+[\6_\m\t_4^\n-\d^\n_\m(\t_1+m^2)]
\frac{\d}{\d\t_3^\n}-\t_2\frac{\d}{\d\t_4^\m}
\Bigg).
\eqan{ward}
Fortunately, the breaking term $\D_\m^{cl}$ is linear in the quantum 
fields and 'insertions' of linear quantum fields are not renormalized 
by quantum corrections. It is given by 
\eqa
\D_\m^{cl} &=& \int_\MM d^2x \left( L^a  \6_\m c^a-\r^a 
\6_\m\f^a-\O^{\n a} \6_\m 
A_\n^a-\e_\mn \r^a \6^\n b^a + \L^{a}\6_\m \l^a\right.
\nonumber \\
 &-& \left.Y^{a} \6_\m X^a -\s^{\n a}\6_\m B_{\n}^a-  
\e_\mn Y^{a}\6^\n d^a+\e_\mn\r^a\t_3^\n\cb^a+\e_\mn\r^a\t_4^\n b^a
\right.\nonumber\\
 &+& \left.\frac{}{}\e_\mn Y^{a}\t_3^\n\lb^a+\e_\mn Y^a \t^\n_4 d^a
\right)\,.
\eqan{break}
Furthermore, the total action $\S^{(0)}$ turns out to be constrained
by: \\
(i) 2 gauge conditions,
\eq
\ba{rcl}
\dfrac{\d \S^{(0)}}{\d b^a} &=& \pa^\m A^a_\m - \t_2 c^a + \t_4^\m A^a_\m, \\
\\[0.1mm]
\dfrac{\d \S^{(0)}}{\d d^a} &=& \pa^\m B^a_\m - \t_2 \l^a + \t^\m_4 B^a_\m.
\ea
\eqn{g.con}
(ii) 2 ghost equations, obtained by commuting the gauge conditions
     with the Slavnov identity \cite{olivier2},
\eq
\ba{rcl}
{\GG}_1\S^{(0)}&=&\dfrac{\d \S^{(0)}}{\d \bar c^a} + (\pa^\m + \t^\m_4) 
\dfrac{\d \S^{(0)}}{\d \O^{a\m}}
+ \t_2 \dfrac{\d \S^{(0)}}{\d L^a} = -(\t_1 + m^2)c^a - \t^\m_3 A^a_\m, \\
\\[0.1mm]
{\GG}_2\S^{(0)}&=&\dfrac{\d \S^{(0)}}{\d \bar \l^a} + (\pa^\m + \t^\m_4) 
\dfrac{\d \S^{(0)}}{\d \s^{a\m}}
+ \t_2 \dfrac{\d \S^{(0)}}{\d \L^a} = -(\t_1 + m^2)\l^a - \t^\m_3 B^a_\m.
\ea
\eqn{a.eq}

\noindent (iii) 2 antighost equations, \\
\eq
\ba{rcl}
\bar{\GG}_1^a \S^{(0)} &=& \D^a_1, \\
\bar{\GG}_2^a \S^{(0)} &=& \D^a_2,
\ea
\eqn{lkiri}
where
\eq
\bar{\GG}^a_1 = \int_\MM d^2x \bigg( \frac{\d }{\d c^a} - f^{abc} \bar c^b 
          \frac{\d}{\d b^c} - f^{abc} \bar \l^b \frac{\d}{\d d^c}
          \bigg).
\eqn{lksdj1}
The corresponding breaking $\D^a_1$ is linear in the quantum fields
\eq
\D^a_1 = \int_\MM d^2x \bigg\lbrace f^{abc} \bigg(- \O^{b\m} A_\m^c + L^b c^c - 
                   \r^b \f^c - \s^{b\m} B^c_\m + \L^b \l^c - Y^b X^c \bigg) 
                   + (\t_1 + m^2) \bar c^a + \t_2 b^a \bigg\rbrace. 
\eqn{lksdj2}
For the second antighost equation we have
\eq
\bar{\GG}_2^a = \int_\MM d^2x \left( \frac{\d}{\d \l^a} - f^{abc} \lb^b
          \frac{\d}{\d b^c} \right),
\eqn{anti2}
such that
\eq
\D^a_2 = -\int_\MM d^2x \Big( f^{abc}(\r^b X^c - \L^b c^c + \s^{b\m}A^c_\m)
         - (\t_1 + m^2)\lb^a - \t_2d^a \Big)\,.
\eqn{j45nm} 
Now, for an arbitrary functional $\G$, depending on the same fields as 
the total action $\S^{(0)}$, 
the corresponding linearized Slavnov operator $\SS_\G$, the
Ward operator for the vector-like supersymmetry  $\VV_\m$, 
and the two antighost operators $\bar{\GG}^a_1$, $\bar{\GG}^a_2$
yield the following nonlinear algebra:
\eq
\ba{rcl}
\SS_\G \SS (\G) &=& 0, \\
\SS_\G (\VV_\m \G - \D_\m^{cl}) + \VV_\m \SS(\G) &=& \PP_\m \G, \\
\left\{ \VV_\m, \VV_\n \right\} \G & = & 0, \\
\SS_\G ( \bar \GG^a_1 \G - \D^a_1 ) + \bar \GG^a_1 \SS(\G)
                 &=& \HH^a \G, \\
\SS_\G ( \bar \GG^a_2 \G - \D^a_2 ) + \bar \GG^a_2 \SS(\G)
                 &=& \KK^a \G, \\
\VV_\m ( \bar \GG^a_1 \G - \D^a_1 ) + \bar \GG^a_1 ( \VV_\m \G
              - \D^{cl}_\m ) &=& 0, \\
\VV_\m ( \bar \GG^a_2 \G - \D^a_2 ) + \bar \GG^a_2 ( \VV_\m \G
              - \D^{cl}_\m ) &=& 0.
\ea
\eqn{mnybcxgkjaj}
$\PP_\m$ is the Ward operator for translations 
\eq
\PP_\m = \int_\MM d^2x \sum_{\Phi_i}\pam \Phi_i \frac{\d}{\d \Phi_i},
\eqn{trans}
where $\Phi_i$ represents collectively all the fields introduced so far.
For the operator $\HH^a$ we have to consider only 
the fields possessing a gauge index (represented by $\Theta^a$) such that
\eq
\HH^a= \int d^2x \sum_\Theta \left(-f^{abc}\Theta^b\frac{\d}
{\d\Theta^c}\right)\,.
\eqn{455ahfg9283}
The operator $\KK^a$ is given by
\eqa
\KK^a &=&  - \int d^2x f^{abc} \Big(X^b\frac{\d}{\d\phi^c} 
+ A_\mu^b\frac{\d}{\d B_\mu} +  
d^b\frac{\d}{\d b^c}+c^b\frac{\d}{\d\l^c}+\lb^b\frac{\d}{\d\cb^c} \non
&+& \s^{\mu b}\frac{\d}{\d\Omega^{\mu c}}+\rho^b\frac{\d}{\d Y^c}+
\L^b\frac{\d}{\d L^c}\Big)\,.
\eqan{kdfj543kj}
Now, it is easy to check that the classical action is invariant under
the symmetries expressed by $\HH^a$ and $\KK^a$, i.e.,
\eq
\HH^a \S^{(0)} = \KK^a \S^{(0)} = 0\,.
\eqn{ja7t843kjsd}
If the functional $\S^{(0)}$ is a solution of the Slavnov identity,
$\SS(\S^{(0)})=0$, of the Ward identities (\ref{invar2}) and 
(\ref{ja7t843kjsd}), and of the
two antighost equations (\ref{lkiri}), then,
from the above nonlinear algebra, we get the following linear algebra: 
\begin{equation}
\label{algebraend}
\begin{array}{rcl}
\left\{ \SS_{\S^{(0)}}, \SS_{\S^{(0)}} \right\} & = & 0,  \\
\left\{ \SS_{\S^{(0)}}, \VV_\m \right\} & = & \PP_\m, \\
\left\{ \VV_\n, \VV_\m \right\} & = & 0, \\
\left\{ \SS_{\S^{(0)}}, \bar \GG^a_1 \right\} & = & \HH^a, \\
\left\{ \SS_{\S^{(0)}}, \bar \GG^a_2 \right\} & = & \KK^a, \\
\left\{ \VV_\m, \bar \GG^a_1 \right\} & = & 0, \\
\left\{ \VV_\m, \bar \GG^a_2 \right\} & = & 0, \\
\end{array}
\end{equation}
So far we have regularized the infrared divergent propagators and analyzed
the symmetries of the model as well as derived the constraints which 
the total action obeys. In the remaining part of the paper we will 
extend our analysis to the quantum level.

\section{Search for counterterms}

We devote this section to the discussion of the stability problem which 
amounts to analyze all possible invariant counterterms for the total action.
In a first step one has to modify the classical action as
\eq
\S'=\S^{(0)}+\D,
\eqn{huiz}
where $\D$ stands for appropriate invariant counterterms. The
total classical action $\S^{(0)}$ is a solution of the
Slavnov identity (\ref{slavnovid}), the vector--like  
supersymmetry WI (\ref{invar2}), the two gauge conditions 
(\ref{g.con}), the two ghost equations (\ref{a.eq}), and the 
two antighost equations (\ref{lkiri}) as well as the two Ward identities
(\ref{ja7t843kjsd}). The perturbation $\D$ is an integrated, 
local and Lorentz invariant polynomial of dimension 2 and vanishing ghost 
number. \\
By studying the stability of the theory we are looking 
for the most general deformation of the classical action such that 
the functional $\S'$ still obeys all the constraints listed 
above. Then the quantity $\D$ is subject to the following set of constraints
\eq
\dfrac{\d \D}{\d b^a} = 0, 
\eqn{cstr1}
\eq
\dfrac{\d \D}{\d d^a} = 0, 
\eqn{cstr2}
\eq
\dfrac{\d \D}{\d \bar c^a} + (\pa^\m + \t^\m_4) \dfrac{\d \D}{\d \O^{a\m}}
+ \t_2 \dfrac{\d \D}{\d L^a} = 0,
\eqn{cstr3}
\eq
\dfrac{\d \D}{\d \bar \l^a} + (\pa^\m + \t^\m_4) \dfrac{\d \D}{\d \s^{a\m}}
+ \t_2 \dfrac{\d \D}{\d \L^a} = 0,
\eqn{cstr4}
\eq 
\SS_{\S^{(0)}} \D = 0,
\eqn{cstr5}
\eq 
\VV_\m \D = 0,
\eqn{cstr6}
\eq
\PP_\m\D=0\,,
\eqn{cstr9}
\eq
\int_\MM d^2x \frac{\d \D}{\d c^a} = 0,
\eqn{cstr7}
\eq
\int_\MM d^2x \frac{\d \D}{\d \l^a} = 0,
\eqn{cstr8}
\eq
\HH^a \D = 0,
\eqn{kjaabf3452nb}
\eq
\KK^a \D = 0.
\eqn{jadhhh345b8f}
The first two equations (\ref{cstr1}) and (\ref{cstr2}) signify that $\D$
is independent of the two Lagrange multiplier fields $b^a$ and $d^a$.
The equation (\ref{cstr3}) implies that $\D$ depends on the fields 
$\O^{a\m}$, $\cb^a$ and $L^a$ only through the two combinations
\eq
\ba{rcl}
\tilde \O^{a\m} &=& \O^{a\m} + \pa^\m \bar c^a - \t^\m_4 \bar c^a, \\
\tilde L^a &=& L^a + \t_2 \bar c^a\,.
\ea
\eqn{comb1}
Correspondingly, we deduce from equation (\ref{cstr4}) that the fields
$\s^{a\m}$, $\lb^a$ and $\L^a$ appear in the expression of $\D$ only
through the two combinations
\eq
\ba{rcl}
\tilde \s^{a\m} &=& \s^{a\m} + \pa^\m \bar \l^a - \t^\m_4 \bar \l^a, \\
\tilde \L^a &=& \L^a + \t_2 \bar \l^a.
\ea
\eqn{comb2}
Finally, the constraints (\ref{cstr5}), (\ref{cstr6}) and (\ref{cstr9})
may be combined in a single Ward-operator $\d$, 
\eq
\d = \SS_{\S^{(0)}} + \x^\m \VV_\m + \ve^\m \PP_\m - \int_\MM d^2x ~
\x^\m \frac{\6 }{\6 \ve^\m}.
\eqn{delta}
It is easy to check the nilpotency of the above defined operator $\d$.
The vectors $\x^\m$ and $\ve^\m$ are constant vectors of ghost numbers
$+2$ and $+1$, respectively. Clearly, we get 
\eq
\d \D = 0,
\eqn{co}
which constitutes a cohomology problem. The nilpotency property of the operator
$\d$ implies immediately that any expression of the form 
$\d \hat \D$ is a solution of (\ref{co}), where $\hat \D$ is a local 
integrated polynomial of dimension 2 and ghost number $-1$. Therefore,
the general solution of (\ref{co}) is of the  form
\eq
\D = \D_c + \d \hat \D\,,
\eqn{mxcipwe345}
where $\D_c$ is the nontrivial solution whereas $\d \hat \D$ is called the
trivial solution. \\
The form of the trivial counterterm is restricted by  dimension and 
ghost number requirements. Since the fields $(\f, X)$ both have 
dimension and ghost number zero, an arbitrary combination of these 
fields may appear many times in the counterterm.
We denote the most general and possible combination of $(\f, X)$ by 
$f_\alpha[\f,X]$, such that
\begin{equation}
\label{falpha}
f_\alpha[\f,X]=\sum_{\{n_i\},\{m_i\}=0}^\infty \b^\a_{n_i,m_i}\Bigg(
\prod_{i=0}^\infty
\f^{n_{i}} X^{m_{i}}\Bigg)\,,
\end{equation}
where $\{n_i\}$ and $\{m_i\}$ are understood as $\{n_0,n_1,\dots\}$ and 
$\{m_0,m_1,\dots\}$, respectively. 
$\b^\a_{n_i,m_i}$ are, of course, constant coefficients to be determined.
Actually, the most general trivial counterterm $\d\hat{\D}$ reads
\begin{eqnarray}\label{counter}
\d \hat{\D} &=& \d \int_\MM d^2 x \hbox{Tr} \left(\frac{}{}
\rho f_1 + 
Y f_2 + 
\t_2 f_3 + 
\tilde{\O}^\n f_4 A_\n f_5+
\e_{\m\n}\tilde{\O}^\m f_6 A^\n f_7+ 
\tilde{\s}^\n f_8 A_\n f_9
\right.\nonumber\\
&+&
\e_{\m\n}\tilde{\s}^\m f_{10} A^\n f_{11} +
\tilde{\O}^\n f_{12} B_\n f_{13} + 
\e_{\m\n}\tilde{\O}^\m f_{14} B^\n f_{15} + 
\tilde{\s}^\n f_{16} B_\n f_{17} 
\nonumber\\ 
&+&
\e_{\m\n}\tilde{\s}^\m f_{18} B^\n f_{19}+ 
(\6^\n \tilde{\O}_\n) f_{20} +
\e_{\m\n}(\6^\m \tilde{\O}^\n) f_{21} +
 (\6^\n \tilde{\s}_\n) f_{22} +
 \e_{\m\n} (\6^\m \tilde{\s}^\n) f_{23} 
\nonumber\\ 
&+&
 \t_4^\n f_{24}\tilde{\O}_\n f_{25} +
 \e_{\m\n}\t_4^\m f_{26}\tilde{\O}^\n f_{27} +
 \t_4^\n f_{28} \tilde{\s}_\n f_{29} +
 \e_{\m\n}\t_4^\m f_{30}\tilde{\s}^\n f_{31} 
\nonumber\\ 
&+&
\left.\frac{}{}
 \tilde{L}f_{32} c f_{33} + 
\tilde{\L}f_{34} c f_{35} +
 \tilde{L}f_{36} \l f_{37} + 
\tilde{\L} f_{38}\l f_{39}
\right) \,.
\end{eqnarray}
In fact the expression (\ref{counter}) may depend on the vector parameters
$\x^\m$ and $\ve^\m$ which are not present in the total action 
(\ref{totaction}).
For this reason we have to arrange for the trivial counterterm to be 
independent\footnote{ Our aim is to renormalize the theory defined by 
(\ref{totaction}) which is independent of $\x^\m$ and $\ve^\m$.}
of these two constant vectors.
In other words, we require that $\hat \D$ has to be invariant under 
the vector--like
supersymmetry and translation Ward operators.
A lengthy and detailed analysis results in the expression
for the trivial counterterm given below:
\begin{eqnarray}
\label{nmxc54nmf9ksd}
\SS_{\S^{(0)}} \CC_t&=& \SS_{\S^{(0)}} \int_\MM d^2 x \hbox{Tr} 
\left(\frac{}{} \b^1\,(\tilde{\O}^\n A_\n+\rho\f-\tilde{L}c)+\b^2\,
(\tilde{\s}^\n A_\n+Y\f-\tilde{\L}c) ~+
\right.\nonumber\\
&+& \left.\b^3\,(\tilde{\O}^\n B_\n+\rho X-\tilde{L}\l)+
\b^4\,(\tilde{\s}^\n B_\n+Y X -\tilde{\L}\l)\right)\,,
\end{eqnarray}
where the $\b^i$, $i=1,...,4$ are arbitrary constants. \\
Now we turn to the computation of the nontrivial counterterms $\D_c$
in (\ref{mxcipwe345}). In a first step we introduce a
filtering operator $\NN$ such that
\eq
\NN = \int_\MM d^2x \Bigg( \sum_f f \frac{\d}{\d f} \Bigg),
\eqn{filter}
where we have assigned to each field homogeneity
degree one. The quantity $f$ in (\ref{filter}) stands for all fields
(including also $\e^\m$ and $\x^\m$). The operator $\NN$ leads to the 
decomposition
of the operator $\d$ as
\eq
\d = \d_0 + \d_1\,.
\eqn{decomp}
The nilpotency of the operator $\d$ implies now that
\eq
\d_0^2 = \lbrace \d_0,\ \d_1 \rbrace = \d_1^2 = 0.
\eqn{d0nilpot}
The operator $\d_0$ does not increase the homogeneity degree, whereas 
$\d_1$ increases the homogeneity degree by one unit. 
Now it is evident that from $\d \D = 0$ we get
\eq
\d_0 \D = 0,
\eqn{co2}
where
\eqac
\d_0 &=& \displaystyle{\int}_\MM \left( d c^a \frac{\d }{\d A^a} + 
         d A^a \frac{\d }
         {\d  \r^a} + d \f^a \frac{\d }{\d \hat \O^a} + 
         d \hat \O^a \frac{\d }{\d \hat L^a} \right) + \non \\
  &+&    \displaystyle{\int}_\MM \left( d \l^a \frac{\d }{\d B^a} + 
         d B^a \frac{\d }
         {\d  \hat Y^a} + d X^a \frac{\d }{\d \hat \s^a} + 
         d \hat \s^a \frac{\d }{\d \hat \L^a} \right) + \non \\
  &+&    \displaystyle{\int}_\MM d^2 x \bigg( 
         - \t_1  \frac{\d }{\d \t_2} + 
         \t^\m_3 \frac{\d }{\d \t^\m_4}  
         - \x^\m \frac{\d }{\d \ve^\m} \bigg)\,.
\eqacn{delta0}
The first two parts of the expression of $\d_0$ are given in terms of forms
where,
\eq
\ba{rcl}
A^a &=& A_\m^a \dxm, \\
\hat \O^a &=& \ve_\mn \Ot^{a\m} \dxn, \\
\hat L^a &=& \frac{1}{2} \ve_\mn \tilde L^a \dxm \dxn, \\
\hat \r^a &=& \frac{1}{2} \ve_\mn \r^a \dxm \dxn, \\
B^a &=& B_\m^a \dxm, \\
\hat \s^a &=& \ve_\mn \tilde \s^{a\m} \dxn, \\
\hat \L^a &=& \frac{1}{2} \ve_\mn \tilde \L^a \dxm \dxn, \\
\hat Y^a &=& \frac{1}{2} \ve_\mn Y^a \dxm \dxn, \\
\ea
\eqn{forms}
and $d$ is the exterior derivative $d = \dxm \pam$.\\
By looking to the expression (\ref{delta0}) one easily recognizes
that the fields $\t_1, \t_2, \t^\m_3, \t^\m_4, \e^\m$ and $\x^\m$
transform as $\d_0$ doublets which means that they are out of the
cohomology of $\d_0$ \cite{brandt}.
In order to solve the cohomology problem (\ref{co2}) we write $\D$ as
an integrated local polynomial: 
\eq
\D=\int_\MM f^0_2,
\eqn{z123z}
such that $f^0_2$ has form degree 2 and ghost number $0$. 
The use of Stoke's theorem, the algebraic Poincare lemma \cite{brandt}
and the anticommutator relation $\lbrace \d_0, ~ d \rbrace =0$ lead to
the set of descent equations
\eq
\ba{rcl}
\d_0 f^0_2 + d f^1_1 &=& 0, \\
\d_0 f^1_1 + d f^2_0 &=& 0, \\
\d_0 f^2_0 &=& 0.
\ea
\eqn{ab.c}
Due to dimension and ghost number requirements the most general expression for 
$f^2_0$ is given by
\eq
f_0^2 = \sum_{n_{ij},m_{ij}, 
                \atop l_{pq},k_{pq}=0}^\infty ~\sum_{j,q=1}^\infty
        \a_{jq} Tr \bigg \lbrack
        ~c \prod_{i=1}^\infty (\f^{n_{ij}} X^{m_{ij}}) 
        ~c \prod_{p=1}^\infty (\f^{l_{pq}} X^{k_{pq}}) \bigg \rbrack ~~+ 
\eqn{solct}
\[
  ~~~  + \sum_{\hat n_{ij},\hat m_{ij}, 
                \atop \hat l_{pq},\hat k_{pq}=0}^\infty ~\sum_{j,q=1}^\infty
        \b_{jq} Tr \bigg \lbrack
        ~c \prod_{i=1}^\infty (\f^{\hat n_{ij}} X^{\hat m_{ij}}) 
        ~\l \prod_{p=1}^\infty (\f^{\hat l_{pq}} X^{\hat k_{pq}}) \bigg 
        \rbrack ~~+ 
\]
\[ 
  ~~~  + \sum_{\bar n_{ij},\bar m_{ij}, 
                \atop \bar l_{pq},\bar k_{pq}=0}^\infty ~\sum_{j,q=1}^\infty
        \g_{jq} Tr \bigg \lbrack
        ~\l \prod_{i=1}^\infty (\f^{\bar n_{ij}} X^{\bar m_{ij}}) 
        ~\l \prod_{p=1}^\infty (\f^{\bar l_{pq}} X^{\bar k_{pq}}) \bigg 
        \rbrack ~~.    
\]
$\a_{jq},\b_{jq}$ and $\g_{jq}$ stand for constant and field independent
coefficients. The upper indices of the fields $\f$ and $X$ are just integer
exponents required by locality.
To solve the descent equations (\ref{ab.c}) we follow the same 
strategy as in \cite{oli}. We define the operator
\eq
\bar \d_0 = \displaystyle{\int}_\MM \Big( 
2 \hat \r \frac{\d}{\d A} + A \frac{\d}{\d c}
+ \hat \O \frac{\d}{\d \f} + 2 \hat L \frac{\d}{\d \hat \O} 
+2 \hat Y \frac{\d}{\d B} +  B \frac{\d}{\d \l} 
+ \hat \s \frac{\d}{\d X} +  2 \hat \L \frac{\d}{\d \hat \s}
\Big),
\eqn{843ghfgz8}
which, when commuted with the operator $\d_0$, gives translations
\eq
\lbrack \d_0 , \bar \d_0 \rbrack = \int_\MM \Big( d \Psi 
\frac{\d}{\d \Psi} \Big)\,.
\eqn{aksdh4kjhasd8}
Recall that we are now working in the space generated by the fields
which belong to the cohomology of $\d_0$. These fields are denoted
by $\Psi$. In other words, the operator
$\d_0$ appearing in equation (\ref{aksdh4kjhasd8}) is restricted
to this space where the $\d_0$ doublets are absent.\\
One can easily show \cite{oli} that the solution of the 
descent equations (\ref{ab.c}) is given by
\eq
f^0_2 = \frac{1}{2} \bar \d_0 \bar \d_0 f^2_0 .
\eqn{solsol234}
$\displaystyle{\int_\MM} f^0_2~$ is then the nontrivial solution of 
the cohomology problem (\ref{co2}).
A direct investigation shows that each monomial in (\ref{solct}) leads 
(after applying on it $\bar{\d_0}^2$) to a polynomial depending on 
the ghost fields $c$ and $\l$. But this is forbidden by the two
constraints (\ref{cstr7}) and (\ref{cstr8}) which are valid
at each homogeneity degree. In other words 
the nontrivial solution of the $\d_0$ cohomology in the space 
constrained by (\ref{cstr7}) and (\ref{cstr8}) is zero.
From this we deduce\footnote{ Recall \cite{brandt} 
\cite{olivier2} that the cohomology of the nilpotent operator $\d$, 
that is the space of all nontrivial solutions of (\ref{co}), is isomorphic 
to a subspace of the cohomology of $\d_0$.}
that the cohomology of the whole operator $\d$ is empty.
So, the most general solution of (\ref{co}) takes the form
\eq
\D = \SS_{\S^{(0)}} \CC_t \, .
\eqn{kafhgi342z8df}
The restriction coming from the
two antighost equations (\ref{cstr7}) and   
(\ref{cstr8}) eventually implies the vanishing of all constant coefficients in 
(\ref{nmxc54nmf9ksd}). \\
Thus we have shown that the constraint system (\ref{cstr1}--\ref{cstr8}) 
forbids any deformations of the classical action. Furthermore, if the 
symmetries of the
model are anomaly free, then the symmetry content of the model at the 
classical level is also valid in the presence of quantum corrections.
The analysis of anomalies is the subject of the next section.   

\section{Search for anomalies}

In order to describe possible breaking of the symmetries which characterize 
the model, one has to apply the quantum action principle (QAP) \cite{olivier2}.
The latter allows to describe symmetry breaking in the following way:
\eq
\d\G=\tilde{\D} \, ,
\eqn{symbr}
where $\G$ is the full vertex functional given by a power series in $\hbar$.
The QAP requires that the breaking $\tilde{\D}$ is a local, integrated, 
Lorentz-invariant
polynomial of dimension 2 and ghost number 1. The nilpotency of $\d$ leads
again to a cohomology problem
\eq
\d\tilde{\D}=0\,,
\eqn{cohpr}
which implies the solution, 
\eq
\tilde\D = \d\tilde{\tilde{\D}}+\AA\,,
\eqn{sllsdkjaekrihtpoe}
where $\AA\not=\d\tilde{\AA}$.
The anomaly candidate $\AA$, as a solution of (\ref{cohpr}), has to 
obey\footnote{ A non trivial statement is that the anomaly candidate 
has to fulfill the antighost equations.
Due to the algebra (\ref{algebraend}) we deduce
\eq
\ba{rcl}
\lbrace \d, \bar \GG^a_1 \rbrace &=& \HH^a, \\
\lbrace \d, \bar \GG^a_2 \rbrace &=& \KK^a.
\ea
\eqn{kfg9g9fido9}
Furthermore, the quantum generating functional of vertex functions 
$\G = \S^{(0)} + \OO (\hbar)$ fulfills the two antighost equations
and the two WI's (\ref{ja7t843kjsd}), see below. 
This fact together with (\ref{symbr}) and (\ref{kfg9g9fido9})
shows that the anomaly candidate must obey the antighost equations.} 
the constraints (\ref{cstr1}) --- (\ref{cstr4}) as well as (\ref{cstr7})
and (\ref{cstr8}). In other words we have to solve the cohomology problem
(\ref{cohpr}) in the same space of functionals as the problem (\ref{co}),
Hence, $\AA$ depends only on the fields: $A^a$, $c^a$, $\r^a$, $\vf^a$,
$\hat \O^a$, $\hat L^a$, $\l^a$, $B^a$, $\hat Y^a$, $X^a$, $\hat \s^a$ and
$\hat \L^a$.
In terms of forms the functional $\AA$ is a local integrated 
polynomial of form degree 2 and ghost number 1
\eq
\AA = \int_\MM f^1_2.
\eqn{lksdfh33}
By using the strategy of the previous section we get the following 
set of descent equations
\eq
\ba{rcl}
\d_0 f^1_2 + d f^2_1 &=& 0, \\
\d_0 f^2_1 + d f^3_0 &=& 0, \\
\d_0 f^3_0 &=& 0.
\ea
\eqn{ab.a}
The last equation in (\ref{ab.a}) has the general solution
\eq
f_0^3 = \sum_{n_{ij},m_{ij},l_{pq}, 
                \atop k_{pq},t_{gy},h_{gy}=0}^\infty ~\sum_{j,q,y=1}^\infty
        \a_{jqy} Tr \bigg \lbrack
        ~c \prod_{i=1}^\infty (\f^{n_{ij}} X^{m_{ij}}) 
        ~c \prod_{p=1}^\infty (\f^{l_{pq}} X^{k_{pq}}) 
        ~c \prod_{g=1}^\infty (\f^{t_{gy}} X^{h_{gy}}) \bigg \rbrack ~~+ 
\eqn{solan}
\[
  ~~~  + \sum_{\hat n_{ij},\hat m_{ij},\hat l_{pq}, 
                \atop \hat k_{pq},\hat t_{gy},\hat h_{gy}=0}^\infty 
        ~\sum_{j,q,y=1}^\infty
        \b_{jqy} Tr \bigg \lbrack
        ~c \prod_{i=1}^\infty (\f^{\hat n_{ij}} X^{\hat m_{ij}}) 
        ~c \prod_{p=1}^\infty (\f^{\hat l_{pq}} X^{\hat k_{pq}}) 
        ~\l \prod_{g=1}^\infty (\f^{\hat t_{gy}} X^{\hat h_{gy}}) 
        \bigg \rbrack ~~+
\]
\[ 
  ~~~  + \sum_{\bar n_{ij},\bar m_{ij},\bar l_{pq}, 
             \atop \bar k_{pq},\bar t_{gy},\bar h_{gy}=0}^\infty 
        ~\sum_{j,q,y=1}^\infty
        \g_{jqy} Tr \bigg \lbrack
        ~c \prod_{i=1}^\infty (\f^{\bar n_{ij}} X^{\bar m_{ij}}) 
        ~\l \prod_{p=1}^\infty (\f^{\bar l_{pq}} X^{\bar k_{pq}}) 
        ~\l \prod_{g=1}^\infty (\f^{\bar t_{gy}} X^{\bar h_{gy}}) \bigg 
        \rbrack ~~+ 
\]
\[
  ~~  +\sum_{\tilde n_{ij},\tilde m_{ij},\tilde l_{pq}, 
                \atop \tilde k_{pq},\tilde t_{gy},\tilde h_{gy}=0}^\infty 
        ~\sum_{j,q,y=1}^\infty
        \pi_{jqy} Tr \bigg \lbrack
        ~\l \prod_{i=1}^\infty (\f^{\tilde n_{ij}} X^{\tilde m_{ij}}) 
        ~\l \prod_{p=1}^\infty (\f^{\tilde l_{pq}} X^{\tilde k_{pq}}) 
        ~\l \prod_{g=1}^\infty (\f^{\tilde t_{gy}} X^{\tilde h_{gy}}) 
        \bigg \rbrack,  
\]
where the quantities $\a_{jqy},\b_{jqy},\g_{jqy}$ and $\pi_{jqy}$ are
constant coefficients.\\
By using the same arguments as before (see the last section) one 
can prove that all constant coefficients appearing in (\ref{solan})
must vanish. Of course, this is due to the constraints (\ref{cstr7})
and (\ref{cstr8}). Let us, for instance consider the special case
where
\eq
f_0^3 = \a Tr(c)^3 + \b Tr(\l)^3 + \g Tr(c^2 \l),
\eqn{bla1}
It leads to the anomaly candidate 
\eq
\AA = Tr \int_\MM \bigg( 3\a(\hat \r c^2 + A^2 c) + 3~\b(\hat Y \l^2 + B^2 \l)
              + \g ( \hat \r \lbrace c,~ \l \rbrace +
               \lbrace c,~ A \rbrace B + A^2 \l + \hat Y c^2 ) \bigg),
\eqn{aaa}
where $\a$, $\b$ and $\g$ are constant coefficients.\\
However, it is easy to verify that this anomaly candidate 
(\ref{aaa}) does not obey the two antighost equations (\ref{cstr7}) and 
(\ref{cstr8}) unless $\a=\b=\g=0$. 
This means in particular that the nontrivial solution of 
(\ref{cohpr}) is zero. Then, 
the Slavnov identity as well as the 
WI for the vector-like supersymmetry transformations 
are anomaly free. Hence, they are valid at the full quantum level.
Concerning the two antighost equations, one can prove their 
validity at the quantum level by simply following the 
arguments of \cite{bps}.
Furthermore, the two gauge conditions, the two ghost 
equations and the two WI's (\ref{ja7t843kjsd}) are also valid 
at the quantum level. This can be proven by
simply using the strategy of \cite{olivier2}.  \\
As a conclusion, the model we analyzed in this paper is anomaly free
and ultraviolet
as well as infrared finite at all orders of perturbation theory.



\begin{thebibliography}{99}
\addcontentsline{toc}{section}{Bibliography}

\bibitem{blau93}M. Blau, G. Thompson, hep-th/9310144.
%
\bibitem{2dgravity}
A.H. Chamseddine and D. Wyler, \np{B340}{90}{595}\\
{\em Phys. Lett.} {\bf B228} (1989) 595.
%
\bibitem{blasi}A. Blasi and N. Maggiore, {\em Class. Quantum Grav. } {\bf 10 } 
(1993) 37.
%
\bibitem{balasin}H. Balasin, W. Kummer, O. Piguet and M. Schweda, {\em 
Mod. Phys. Lett. } {\bf A37} (1994) 3487-3493.
%
\bibitem{olivier2} O. Piguet and S.P. Sorella,
{\it Algebraic Renormalization},
Lecture Notes in Physics, Vol. m 28, Springer Verlag, 1995.\\
A. Boresch, S. Emery, O. Moritsch, M. Schweda, T. Sommer, H. Zerrouki, 
{\em Applications of noncovariant gauges in the algebraic renormalization 
procedure }, to be published. 
%
\bibitem{bps}A. Blasi, O. Piguet and S.P. Sorella, {\em Nucl. Phys. }
{\bf B356} (1991) 154.
%
\bibitem{brandt} F. Brandt, N. Dragon and M. Kreuzer,
\pl{B231}{89}{263} \\
\np{B332}{90}{224} \\
{\em Nucl. Phys. }{\bf B340} (1990) 187.
%
\bibitem{oli} O. Piguet, {\it On the Role of Vector Supersymmetry
in Topological Field Theory}, preprint UVGA -- DPT 1995/02-880;
hep-th/9502033.
%
\bibitem{brs}C.M. Becchi, A. Rouet and R. Stora, {\em Comm. Math. Phys. } 
{\bf 42} (1975) 127; {\em Ann. Phys. (N.Y.)} {\bf 98} (1976) 287.
%
\bibitem{brt}D. Birmingham, M. Rakowski and G. Thompson,
{\em Nucl. Phys. } {\bf B329} (1990) 83. 
%
\bibitem{tft}D. Birmingham, M. Blau, M. Rakowski and G. Thompson,
{\em Phys. Rep. } {\bf 209} (1991) 129. 


\end{thebibliography}
\end{document}